# Search Engine Optimization Techniques Practiced in Organizations: A Study of Four Organizations

Muhammad Akram, Imran Sohail, Sikandar Hayat, M. Imran Shafi, and Umer Saeed

**Abstract**—Web spammers used Search Engine Optimization (SEO) techniques to increase search-ranking of web sites. In this paper we have study the essentials SEO techniques, such as; directory submission, keyword generation and link exchanges. The impact of SEO techniques can be applied as marketing technique and to get top listing in major search engines like Google, Yahoo, and MSN. Our study focuses on these techniques from four different companies' perspectives of United Kingdom and Pakistan. According to the these companies, these techniques are low cost and high impacts in profit, because mostly customers focus on major search engine to find different products on internet, so SEO technique provides best opportunity to grow their business. This paper also describes the pros and cons of using these searh engine optimization techniques in above four companies. We have concluded that these techniques are essential to increase their business profit and minimize their marketing cost.

———————— ◆ ————————

## 1 INTRODUCTION

INTERNET spammer's (or web spammer) used search engine optimization techniques to boost their websites of low ranking web sites into highest level search rankings. Mostly used search engine optimizations techniques include material keyword generation, directory submission and link exchanges etc [1]. Two ways of internet marketing are associated with web search engines: Paid placement and search engine optimization. The paid placement is tackled search engine from sponsored or paid results. The internet search engine shows results with quality but these search engines placed charges to boost their web sites. They calculate charges by placing keywords or clicking on those keywords means cost per click. Search engine optimization is another way to optimizing web sites with free of cost. Search engine optimization is going to very popular. According to the Bo Xing and Zhangxi Lin [4] with two conditions cost of keywords has increase 19% and natural results are seen intention and unbiased than paid or sponsored results [4]. Unlimited keywords are placed in internet search engine daily and internet users wait for high quality of results and they retrieve them [2].To achieve high rankings in the search engines link manipulation is important and another way to get these tasks in web search engines. The exchanging links between other websites is the way to draw link exchanges or reciprocal links.

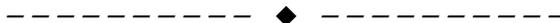

- *Muhammad Akram is Lecturer in College of Computer Science & Information Systems, Najran University, Saudi Arabia.*
- *Imran Sohail is Assistant professor in Fatima Jinnah Woman University, Pakistan.*
- *Sikandar Hayat is Student Administrator in Blekinge Institute of Technology, Sweden.*
- *Muhammad Imran Shafi is SEO of SamSoft software-house and visiting Lecturer in Punjab University, Pakistan.*

There are many ways to do link exchange with webmaster. The one is that to show interest in link exchange on web pages. The second is that to send email to other web owners as request for link exchange. The webmaster also send request for link exchange on many discussion forums between specific categories of link exchanges. There are some webmaster shows interests that they will agree on good ranking website to exchange links between related categories [3]. The directories are used to promote websites and improve search engine ranking. The directory submission presents good opportunities when you place link with titles for submission [5]. Directories are too big and they require assured page rank for proper indexing pages. The directory submission plays a vital role to build links and boost web sites with high rankings [6].

## 2 RELATED WORK

According to Bo Xing and Zhangxi Lin [4] internet search engine have many features for information retrieval vehicles and search engine marketing. Internet search engines are ease and freely to use. Sen. addressed [4] the impact of search engine marketing. There are two choices to boost their websites either they select paid placement or Search Engine Optimization (SEO). But the mostly people selected SEO because it is not costly as compared to paid placement.  The directory submission is great approach to create links and increase your web rankings [6]. Directories another importance is that they provide links and data to other databases and also other web search engines [5]. Manual submission is the best technique because two things are different like directory and target users. So that is the reason manual directory submission plays key role for SEO [7].

Keywords are used as a single word or short phrase in the search engines to extract related information. Keyword is very important for internet search engines as well as search engine marketing. Because it is used as a tool to ranking websites in many search engines [8]. The keyword technique is newly approach in SEO. Different types of keyword are used like: query log, advertiser log mining, proximity searches and Meta tag crawlers [8]. But search engines are used query log mining way to generate keywords as well as Meta tag are used in SEO.

The suggestions of link exchanges in internet search engines are used firstly in 1997 [9]. Link exchange approach is used to permit single websites to exchange links with other related websites in the same category with high page rank. Link exchange has been seen fundamental techniques in SEO [10], internet search engines increasingly popular with index links and keywords but links exchanges are consider as objective, democratic and machine readable symbol for web ranking.

## 3 RESEARCH METHODOLOGY

### 3.1 Research Questions
The study consists on analysis and comparison of search engine optimization techniques within the desired organizations. We prepared the following research questions:
RQ1: What are the most important Search Engine Optimization Techniques exercised in dissimilar companies in United Kingdom and Pakistan?
RQ2: What are the pros and cons of Search Engine Optimization Techniques exercised in companies of United Kingdom and Pakistan? RQ3: What are the impacts of Search Engine Optimization Techniques exercised in Pakistan and United Kingdom based internet companies?

### 3.2 Research Method
Qualitative research method will be used to explore above mentioned research questions in this study. The qualitative research methodology according to Creswell [11] is used to formulate data from open ended observations, interviews through telephone and via messenger. Interviews are conducted in terms of data collection from concerned persons that are working in the related search engine optimization companies. Literatures, journals and related articles also discussed for classifying particular search engine optimization techniques. We have prepared questionnaires for conducting interview for data collection. We are analyzing the search engine optimization techniques with the help of questionnaires to perform interview. The Qualitative approach contents are discussed in the following study as:

1. Literature Survey For SEO Techniques
The available SEO techniques in literatures as well as books are presented in Section 4.

2. Data Collection
The collection of data in forms of texts and tables are described in section 5.

3. Data Analysis
Data analysis is presented in section 6. This analysis is based on information that is collected from interview of United Kingdom and Pakistani SEO related companies.

4. Data Validity
The data validity is presented in section 7.

## 4 SEARCH ENGINE OPTIMIZATION TECHNIQUES

### 4.1 Directory Submission
Ricky Mondal says that directory submission [12] is one of the important techniques in SEO to create incoming links to a website through related page and category. A website is created and need to be rank to get good business results. Manually submission to directories is the best approach to rank your website. Internet directory is the platform on World Wide Web for information and links of many websites. Many directories are providing free service to website in directory. Directory submission required some information regarding website. The information like: URL, title, description, keywords, category, and email are required to submit website in directories. To submit website in directories can produce web traffic for your website. This assist you to promote your business needs. The directory submission is used as SEO technique to promote your business [12].

### 4.2 Keyword Generation
Any search engine optimization method used keywords generation process. The keywords are necessary and most important part of SEO. Because all internet search engine requires some words to elaborate information based on these words. The keywords are must be related to your business. Because related keywords boost website in short span of time [13]. According to Terry Fung [14] there are many online tools available to generate keywords relevant your needs like: Word tracker, Yahoo keyword selector tool, Google Ad words keyword tool and Thesaurus etc. By using these tools just put one word related your website like gamming. These tools provide you huge keywords relevant your website. But only keywords are not providing assurance to popularity of website.

### 4.3 Link Exchanges
According to Mike Barus [15] when going to start your business on internet you need reciprocal link or link exchange with other websites. The link exchange is the method in SEO to place link on other websites and other websites place links on your websites means vice versa. There are many types of link exchanges are used like: illustrate interest directly on web pages and other is that send email or discussion forums to show interest for link exchanges. Only related website but with good page rank websites are required to build reciprocal links [16]. For generation reciprocal links requires website title, description, keywords, URL, and email address for both parties [17]. Link exchange boosts website traffic from user who clicks on your website link as well as they increase ranking in web search engine. The disadvantage against reciprocal link is that some sites do frauds [18]. Sharing of all information including customers is the main disadvantage [19].

## 5 DATA COLLECTION

### 5.1 Interview Preparation

Interview method is used as data collection for the related information of search engine optimization techniques from internet organizations. Interview provides face to face, one on one, in person interview and by telephonic interview are used [11].

Semi structured method [20] of interview is performed for the collection of data related search engine optimization techniques from different search engine organizations. We performed open ended questionnaires. Benefit of open ended questions is that some question occurs later so that these questions can be asked any time within interview. Difficult questions can be with no trouble explained during interview. We fix criteria for the selection of interviewees which are:
- Interviewees must be engage in search engine optimizations techniques.
- Interviewees must have related experience and roles should be same as required in SEO.

We prepared some questionnaires with the help of literature survey and some past experience in related field for the purpose of conducting interview. Then finalize all questions and started to contact persons in United Kingdom and Pakistani internet industry. After that we selected time for interview with interviewees and then conduct interview and its takes time 40-60 minutes. All information related SEO were extracted during interview and further clarified answers with interviewees.

### 5.2 Interview Outcomes

This section provides all information that is collected during interview. All information regarding search engine optimization techniques are presented in the form of tables and textual.

This paper contains the study of four organizations of two countries; Pakistan and United Kingdom. Due to some confidential factors we did not presents organizations names. All information regarding organizations is taken from websites as well from interviewees.

- Company A

This company was established in Pakistan 1998 as a global consulting and IT facilities organization specializing in the field: business process management (SEO), software product engineering. Their aim to assist customers and refining business process and reducing cost, improving services and returns. The contacted person is working in this company as SEO specialist from last 4 years. He is specialist in directory submission, link exchanges and keyword generation in SEO techniques.

- Company B

This company provides services as search engine optimization or internet marketing. Also provides search engine marketing strategies to increase web rankings. Their offices located in America and Pakistan. But we focus on Pakistan office. The contacted person working in this organization as SEO teams lead. He has experience more than five years on SEO.

- Company X

This company was established in June 2006. They are SEO specialist in the United Kingdom clients. They focus on top ranking in the Google natural rankings for their clients. The contacted person is working in this company as a SEO manager. He has three years experience in all SEO techniques.

- Company Y

This is SEO specialist based company in United Kingdom. They are offering website optimization, link exchanges and internet marketing. The contacted person of this company is working as a project assistant in SEO department. He has two years experience related SEO.

### 5.3 Most Important SEO Techniques Exercised In Companies

Search Engine Optimization techniques exercised in internet companies are illustrated here. This test present dummy companies names, their countries and which techniques they used. According to contacted persons of both countries, these SEO techniques are important as; Mostly people consult good real directories to find relevant sources. SEO techniques provide valuable votes for website rankings. SEO techniques represents websites themes and important for search engines. The users only commit searches against specific keywords. By using SEO techniques page rank of websites can be increased. SEO techniques like; directory submission, keyword generation and link exchanges are very important in terms of web rankings, generate revenue and boosted their business through web search engines. Aforementioned techniques are extensively used in selected companies.

#### 5.3.1 Pros and Cons of SEO Techniques

The pros and cons of SEO techniques exercised in companies are presented as:

- Directory Submission

The pros of directory submission in Company A are described as; they get maximum links from directories and boosted their web site page rank. The cons of directory submission as SEO technique are illustrate as; they observed that directory submission is time consuming technique and need extra effort to compile.

The pros of directory submission in Company B are illustrates as; the directory submission provides structured listing as well as one place to find all information in web search engines. The cons of directory submission in Company B are presented as; some fake directories are harmful and difficult to maintain.

The pros of directory submission in Company X are; mostly directories provide permanent link and increased traffic to web sites. The cons of directory submission in Company X are; Directory submission is time consuming and tedious technique.

The pros of directory submission in Company Y are presented as; by using directory submission technique page rank will be improve and mostly directory submission services are free. The cons of directory submission in Company Y are presented as; free listing takes time.

- Keyword Generation

The pros and cons of keyword generation in Company A are described as; keyword are first and most important part of every website. The keywords generation is build to attract visitor. Some abbreviation will not provide same information as required and over filing keywords is the bad idea.

The pros and cons of keywords generation in Company B are showed as; they provide proper website representation. But the selection of keywords is critical and also they avoid spamming.

The pros and cons of keyword generation in Company X are illustrates as; targeting and unique keyword generation plays vital role to increase web ranking. Extra effort is required for generation keywords.

The pros and cons of keyword generation in Company Y are; they provides theme of websites and searches against keywords are taken. But pros of keywords generation is that experienced staff required.

- Link Exchanges

The pros and cons of link exchanges in Company A are presented as; sharing traffic provides better visibility of business goals. This technique is time consuming and sometimes costly.

The pros and cons of link exchanges in Company B illustrate as; they enhance traffic and votes for website ranking as well as increase page rank. But difficult to find good link partner and needs relevant link.

The pros and cons of link exchanges in Company X described as; this technique provide good revenue from web and this the assured way to getting high ranking in search engine. But link exchanges are time consuming technique as well as require related website on which links placed. The pros and cons of link exchanges in Company Y are presented as; this technique promotes online business but quality in ranking takes time.

*5.3.2 Impact of SEO Techniques*

The impacts of search engine optimization techniques according to companies of both countries are presented as:

The impacts of search engine optimization techniques in Company A are illustrates as; they are excellent and helps them to get good positions in major search engines as well gaining good profits through SEO techniques.

The impacts of SEO techniques according to company B are; they are applying as a marketing technique and get top listing in major search engines like Google, Yahoo and MSN. According to the organization these techniques are low cost and high impacts in profits. All customers focusing on major search engines to finding products so SEO techniques provide best opportunities to grow business.

The search engine optimization techniques impacts are according to United Kingdom based Company X as; SEO techniques are the way on which number of visitors are increased for websites through high ranking in the search engines because mostly people goes to the search engines if website rank well then more visitors comes and profits ratio will be high.

As according to Company Y impacts of SEO techniques in their business are as; these SEO techniques provides them best opportunities in internet exposure , affordable techniques for their business and boost business efforts in terms of marketing cost, time and profits.

## 6 DATA ANALYSIS

Data analysis section provides different approaches that satisfy the results from companies in the form on interview outcomes and research questions. The connection of interview questionnaires with research questions explores different ways of analysis.

- RQ1 with Data Analysis

*RQ1: What are the most important SEO Techniques exercised in dissimilar companies in United Kingdom and Pakistan?*

The answer of this research question is provided in interview outcomes section. We analysed that all companies of both countries are using SEO techniques in their organizations like:

Directory submission, keyword generation and link exchanges etc. These all techniques are very important for their business profits, marketing cost and boosted rankings in major search engines to catch more customers. All these SEO techniques are most important because directories provide relevant sources from search engine and provide valuable rankings. Hence directory submission is important technique for SEO. Keywords are used to search the related material from the search engines, without using keyword generation it is not possible to promote web ranking in the search engines. So keyword generation is also important part of SEO. The link exchanges or reciprocal linking is also important technique for SEO. Hence overall analysis shows that all these SEO techniques are most important and necessary techniques because these techniques are most important and basic techniques of SEO in all four companies of both countries.

- RQ2 with Data Analysis

*RQ2: What are the pros and cons of Search Engine Optimization Techniques exercised in United Kingdom and Pakistan?*

We asked questions relating pros and cons of search engine optimization techniques. We analysed that both countries working on same as well as all techniques of SEO. We further analysed that these techniques are essentials for their business profits and minimize their marketing cost. Directory submission provides structured listing in search engines for information retrieval, improve page rank etc. On the other hand the con of this technique is; time consuming because takes time to listing in the directories. But these cons are not critical. The next technique is keyword generation that is also important because every search engine extract information against keywords. Keyword provides proper website representation and unique keywords play vital role to improving ranking in the search engines. Some companies hire experienced peoples for keyword generation. The link exchanges are also necessary technique of SEO. Link exchanges enhance web traffic and increase page rank. This also takes times to links in other websites. By using these techniques achievements of business goals are done. Hence overall analysis is that all SEO techniques that are both country companies using are basic and necessary techniques.

- RQ3 with Data Analysis

*RQ3:What are the impacts of Search Engine Optimization Techniques exercised in Pakistan and United Kingdom Based Internet Companies*

SEO has put good impacts on these companies. SEO techniques help to generate high traffic for increasing sale of business. These SEO techniques help to catch customers as well as provide improvements for quality and growth of website. These SEO techniques are affordable and minimizing cost, time and increase profits.

## 7 DATA VALIDITY

William Trochim says that [21], the validity of the interview outcomes are judged according to some principles that are; Credibility, Transferability, Dependability and Conformability.

- Credibility

To proof research credible literature survey is done for extracting information related SEO techniques. The questionnaire was prepared to perform interviews for data collection. All the four companies of both countries were contacted for interviews. To obtain credible information telephonic interviews as well as messengers are used.

- Transferability

Transferability is directly connected with the results or outcomes from interviews on SEO techniques. These research results will be helpful for internet companies to improve their SEO techniques according to their business needs. The one threat that interviewee belongs two different countries. They both have different education level and cultural background. The interviews with these interviewees helped us to collect information from telephonic interviews style and messengers.

- Dependability

We contacted interviewees for taking time for the interviews. We conduct interviews by telephone and messengers. We observed that some interviewees did not know the terminologies that are performed in their organizations. So that is the reason we selected same types of companies for the results. Some companies are not using SEO techniques this may be effect on interview outcomes. All companies are used SEO techniques which are investigation occurred, hence there is no validity threat regarding organizations.

- Conformability

Semi structured interviews was used to conduct interviews. An open ended question was asked to collect related information. We purified all questionnaires for confirmation that was build main focus on our research. There was no threat occurs during telephonic interviews and via messenger interviews.

## 8 CONCLUSION & FUTURE WORK

This paper presents an interview study of SEO techniques of four companies of United Kingdom and Pakistan. This study focus on SEO techniques, which are exercised in companies like; directory submission, keyword generation and link exchanges. The pros and cons of each SEO techniques exercised according to companies are discussed and analyzed.

The impact of SEO techniques was analysed according to companies. These techniques help to generate high traffic for increasing sales of business, to catch customers as well as provide improvement for quality and growth of websites. Futher these techniques are affordable and minimizing cost of marketing and at the same time it increase the profit. Hence overall analysis shows that all these SEO techniques are most important and necessary techniques in all four companies of both countries to promote their business.

During our research we came to know that directory submission and link exchange takes time for listing in the directories and to link in other websites respectively. So this can be our future work to findout the ways how we can decrease this time in directory submission and link exchange.

## 9 APPENDIX

Semi Structured Interview Questionnaire
1. Do you use Search Engine Optimization Techniques in your company?
2. Which Search Engine Optimization Techniques you follow in your projects?
- Directory Submission
  a) Why directory submission is important for your company as Search Engine Optimization Techniques?
  b) Can you provide two pros and two cons of directory submission as SEO techniques?
  c) What are the Impacts of using directory submission as SEO techniques?

- Link Exchanges
  a) Why link exchanges are important for your company as Search Engine Optimization Techniques?
  b) Can you provide two pros and two cons of Link Exchanges as SEO techniques?
  c) What are the Impacts of using Link Exchanges as SEO techniques?
- Keyword Generation
  a) Why Keyword Generations are important for your company as Search Engine Optimization Techniques?
  b) Can you provide two pros and two cons of Keyword Generation as SEO techniques?
  c) What are the Impacts of using Keyword Generation as SEO techniques?

**Muhammad Akram** obtained Master of Science in Computer Science from Blekinge Institute of Technology Sweden in 2008, Masters in Computer Science (MCS) in 2005, and BSc in 1999 from University of Azad Jammu & Kashmir, Pakistan. Currently he is working as a Lecturer in College of Computer Science and Information System, Najran University, Saudi Arabia.

**Imran Sohail** completed his Master of Science degree from Sweden in 2009 and B.E in 2000. Currently he is working as Assistant professor in Fatima Jinnah University. Before joining FJWU, he was working as a web developer in Eriscsson Stockholm, Sweden. He has vast experience of programming and has developed number of applications.

**Sikandar Hayat** has obtained Masters in Electrical Engineering with emphasis in telecommunication from Blekinge Institute of Technology Sweden in 2009, Master in Information Technology and BSc from University of Azad Jammu & Kashmir Pakistan. Currently he is working as Student Administrator & Project Assistant in IT department of Blekinge Institute of Technology, Sweden.

**Muhammad Imran Shafi** completed Master of Science in Computer Science from Blekinge Institute of Technology Sweden in 2008, MSc in Computer Science and BSc from Punjab University Pakistan. Currently he is SEO of SamSoft (software house) in Lahore and visiting Lecturer in Punjab University Pakistan.

**Umer Saeed** completed MSc in Computer Science from Blekinge Institute of Technology Sweden in 2009, and BS in Computer Science from Punjab Institute of Computer Science Lahore Pakistan.